# Symplectic integration without roundoff error

**David J.D. Earn**

*Current address:*

Racah Institute of Physics
The Hebrew University of Jerusalem
Jerusalem 91904, Israel

E-mail: earn@astro.huji.ac.il

# Symplectic Integration Without Roundoff Error

David J.D. EARN [1 2]

[1] Institute of Astronomy, Madingley Road, Cambridge CB3 0HA, UK
[2] Physics Department, Weizmann Institute, Rehovot 76100, Israel

## 1. Introduction

At every moment dynamical systems are being evolved with the aid of numerical integration algorithms. Unfortunately, it is not always easy to determine whether even the qualitative features of differential equations are well-preserved by numerical "solutions". This is especially true for systems that display chaotic behaviour, and particularly so if the apparent chaos is weak and slow as in the evolution of the planetary orbits (see the recent review by Duncan & Quinn 1993).

A numerical integration method is said to be order $k$ if the theoretical or *truncation error* after one time-step $\Delta t$ is $O(\Delta t^{k+1})$. Truncation error is an unavoidable consequence of taking a finite time-step. Additionally, *roundoff error* occurs if machine arithmetic is not done exactly; this is normally the case if floating-point numbers are used and it introduces unwanted noise to integrations.

With a time-step $\Delta t$ and floating-point numbers with $P$ significant binary digits we may estimate the relative importance of the two types of error with what we shall call the *error ratio*,

$$\mathcal{R}(P, \Delta t, k) = \frac{2^{-P}}{\Delta t^{k+1}} \ . \tag{1}$$

Strictly, the total roundoff error depends in a machine-specific way on the sequence of arithmetic operations required at each step ($2^{-P}$ is a lower limit for quantities of order unity) and the truncation error is of the form $C\Delta t^{k+1}$ only in the limit $\Delta t \to 0$ (we have ignored the constant $C$). Nevertheless, Eq. (1) can be used as a rough guide: roundoff may be unimportant if $\mathcal{R} \ll 1$ but dominates the computational error if $\mathcal{R} \gg 1$. To achieve a given level of truncation error, it is much more efficient to use a modest time-step with a high order method than a tiny time-step with a low order method, so in practice $\mathcal{R} \ll 1$ in low order implementations and $\mathcal{R} \gg 1$ in high order implementations (such as those used in most solar system integrations).

Integration errors are usually quantified by monitoring the evolution of constants of the motion, such as the total energy in a Hamiltonian system. For example, the maximum length of solar system integrations has been guided by an energy error tolerance. Quinn & Tremaine (1990) showed that energy errors could be greatly reduced by carefully avoiding (mainly) the bias of computers to round *up* in all floating-point additions. This trick does not eliminate roundoff error but substantially reduces its influence, at a cost of about a factor of 2 in computing speed.

The magnitude of the errors is important, but so is the character of the errors in long-term studies that aim to determine qualitative behaviour. In the case of the solar system, we would like to distinguish regular from weakly chaotic motion so we would like some reassurance that the numerical errors (principally roundoff) do not significantly alter the qualitative features of the orbits.

The simplest way to start in analyzing roundoff error is to consider a system in which the only error is roundoff. This is the case for *maps*. Earn & Tremaine (1991, 1992; hereafter ET) showed that roundoff errors cause artificial drifting across invariant curves in Hamiltonian maps and can even lead to confusion between regularity and chaos. The elimination of roundoff error in area-preserving maps of the plane was first considered by Rannou (1974). ET showed that Rannou's method can be applied to Hamiltonian maps of arbitrary dimension. To eliminate roundoff error, a given map is slightly perturbed so that it maps a lattice of points to itself and can be iterated exactly on a computer. As shown in ET (and Scovel 1991) these *lattice maps* have the same mathematical structure as the original maps, i.e., they are Hamiltonian (unlike the maps induced by applying floating-point arithmetic to the original formulae).

Lattice maps are known to reduce qualitative errors (e.g., ET). Here, the lattice approach is applied to a fourth-order integration algorithm and it is found that *quantitative* errors (in the conservation of integrals) can also be significantly reduced, even in the short term.

## 2. Symplectic Integration and Lattice Maps

Most numerical integration algorithms are not designed specifically for Hamiltonian systems and do not respect their characteristic properties, which include the preservation of phase space volume with time (Liouville's theorem). This can lead to spurious damping or excitation. Methods that do preserve all the Hamiltonian properties, i.e., for which the time-forward map is symplectic, are called *symplectic integration algorithms* or SIAs (e.g., Channell & Scovel 1990; Sanz-Serna 1992; Yoshida 1993, and references therein).

Since we are mainly interested here in the gravitational $N$-body problem, we restrict attention to Hamiltonians in potential form, i.e., which can be written $H = \frac{1}{2}\mathbf{v}^2 + U(\mathbf{x})$. A variety of useful SIAs for such Hamiltonians can be derived from two simple *symplectic shears*,

$$S_{\mathbf{x}}(t)\begin{pmatrix}\mathbf{x}\\ \mathbf{v}\end{pmatrix} = \begin{pmatrix}\mathbf{x}+t\mathbf{v}\\ \mathbf{v}\end{pmatrix}, \qquad S_{\mathbf{v}}(t)\begin{pmatrix}\mathbf{x}\\ \mathbf{v}\end{pmatrix} = \begin{pmatrix}\mathbf{x}\\ \mathbf{v}-t\frac{\partial U}{\partial \mathbf{x}}(\mathbf{x})\end{pmatrix}. \tag{2}$$

For example, the (first order) leapfrog scheme is

$$S_{\mathbf{x}}(\Delta t) \circ S_{\mathbf{v}}(\Delta t)\ . \tag{3}$$

The (second order) time-centered leapfrog scheme is

$$S_{\mathbf{x}}(\tfrac{1}{2}\Delta t) \circ S_{\mathbf{v}}(\Delta t) \circ S_{\mathbf{x}}(\tfrac{1}{2}\Delta t)\ . \tag{4}$$

A fourth order SIA is given by

$$S_{\mathbf{x}}(a\Delta t)\circ S_{\mathbf{v}}(b\Delta t)\circ S_{\mathbf{x}}(c\Delta t)\circ S_{\mathbf{v}}(d\Delta t)\circ S_{\mathbf{x}}(c\Delta t)\circ S_{\mathbf{v}}(b\Delta t)\circ S_{\mathbf{x}}(a\Delta t)\ , \tag{5}$$

where $a = 1/(2(2-\beta))$, $b = 1/(2-\beta)$, $c = (1-\beta)a$, $d = -\beta c$, and $\beta = 2^{1/3}$. This method, which we shall call `Ruth_4`, is the fourth-order member in the class of algorithms introduced by Ruth (1983). (All these schemes are immediately generalizable to separable Hamiltonians $H = K(\mathbf{p}) + U(\mathbf{q})$.) `Ruth_4` was discovered by Ronald Ruth and first published by Forest & Ruth (1990).

Although all integrators derived from Eqs.(2) are symplectic in theory, they are not symplectic if implemented using finite-precision arithmetic. This problem can be overcome by replacing the shears $S_{\mathbf{x}}$ and $S_{\mathbf{v}}$ with *lattice shears*,

$$\tilde{S}_{\mathbf{x}}(t)\begin{pmatrix}m\mathbf{x}\\ m\mathbf{v}\end{pmatrix} = \begin{pmatrix}m\mathbf{x}+[tm\mathbf{v}]\\ m\mathbf{v}\end{pmatrix}, \quad \tilde{S}_{\mathbf{v}}(t)\begin{pmatrix}m\mathbf{x}\\ m\mathbf{v}\end{pmatrix} = \begin{pmatrix}m\mathbf{x}\\ m\mathbf{v}-\left[tm\frac{\partial U}{\partial \mathbf{x}}(\mathbf{x})\right]\end{pmatrix}, \tag{6}$$

where $m$ is a (large) constant integer and $[\cdot]$ denotes the nearest point on an integer lattice in phase space. Particles on lattice points are mapped to lattice points by $\tilde{S}_{\mathbf{x}}$ and $\tilde{S}_{\mathbf{v}}$ since integer additions can be done exactly. As shown in ET, lattice shears are symplectic so lattice leapfrog and lattice `Ruth_4` are *exactly symplectic in practice* despite the use of finite-precision arithmetic. Using a lattice SIA is equivalent to evolving the exact solution of a problem with a Hamiltonian that is slightly different from the original.

The leapfrog methods are not likely to benefit much from the lattice approach for two reasons: (i) For practical time-steps $\mathcal{R} \ll 1$, even if only single-precision (four byte) floating-point arithmetic is used. (ii) To avoid loss of precision in the force and velocity components when using a lattice map we must have $m\Delta t \geq 2^P$ (see Eqs.(6)). Therefore, since computers provide integers within finite limits only, there is always a minimum time-step permissible in a lattice SIA, *independent of the order of the method.* For low order methods such as leapfrog the minimum $\Delta t$ is typically too large to obtain acceptably small truncation error.

For a fourth-order integrator the role of roundoff error is significant. We concentrate on the `Ruth_4` algorithm here.

## 3. Orbital Elements

The gravitational two-body problem is fully integrable and is equivalent to the motion of a single particle in a Kepler potential $(-1/r)$. There are five isolating integrals of the motion (e.g., Binney & Tremaine 1987). In celestial mechanics the integrals are normally expressed as geometric quantities known as the *orbital elements* (e.g., Brouwer & Clemence 1961).

The shape of the (elliptical) orbit is determined by its *semi-major axis* $a$ and *eccentricity* $e$. The current phase of the orbit is given by its *mean anomaly* $\ell(t)$, and $\ell(0) \equiv \ell_0$ is a constant of the motion (not an isolating integral). The orbit's orientation is determined by its *argument of pericentre* $\omega$ (the angle in the orbit plane from ascending node to pericentre), *longitude of ascending node* $\Omega$, and inclination $I$. In terms of the orbital elements, the energy is $E = -GM/2a$ and the magnitude of angular momentum is $h = \sqrt{GMa(1-e^2)}$, where $M$ is the total mass ("sun" plus "planet").

In a perfect integration (no truncation or roundoff error) all the orbital elements would be exactly conserved.

## 4. Sample Integrations

`Ruth_4` has been tested with the gravitational two-body problem by Kinoshita, Yoshida & Nakai (1991, hereafter KYN). To facilitate comparisons, we use the same initial conditions and time-step as KYN in all the tests reported here, namely $a = 1$, $e = 0.1$, $\ell_0 = \omega = \Omega = I = 0.349\,\mathrm{rad} \simeq 20°$ and $\Delta t = 0.01 \simeq 1/628$ orbital periods $(T = 2\pi\sqrt{a^3/GM})$. Units are specified by $G = 1$ and $M = 1$.

Fig. 1 shows the evolution of the six orbital elements for ten orbital periods of the two-body problem. Integrations were conducted with ordinary `Ruth_4` (dotted curves) and lattice `Ruth_4` (solid curves) each employing eight byte (64 bit) arithmetic (in the top four panels the dotted and solid curves lie on top of one another). With a lattice size of $m = 2^{62}$ the minimum permissible time-step is $2^P/m \simeq 0.002$ so $\Delta t = 0.01$ does not degrade the force or velocity components. Put another way, with $P = 53$ and $\Delta t = 0.01$ we require $m \geq 2^{53}/0.01 \sim 2^{60}$ to prevent loss of precision, so $m = 2^{62}$ is large enough. However, the error ratio is $\mathcal{R}(53, 0.01, 4) \simeq 10^{-6}$ so the lattice approach is not expected to improve the integration noticeably. In Fig. 1 there is no apparent difference in the evolution of the first four elements but there is a very clear reduction of errors in $\Omega$ and $I$ in the lattice integration. The explanation is that the `Ruth_4` integrator exactly conserves the angular momentum vector (cf. KYN) so the errors in $\Omega$ and $I$ are entirely due to roundoff despite the fact that $\mathcal{R} \ll 1$. The errors in the other elements are dominated by truncation error (on this timescale at least).

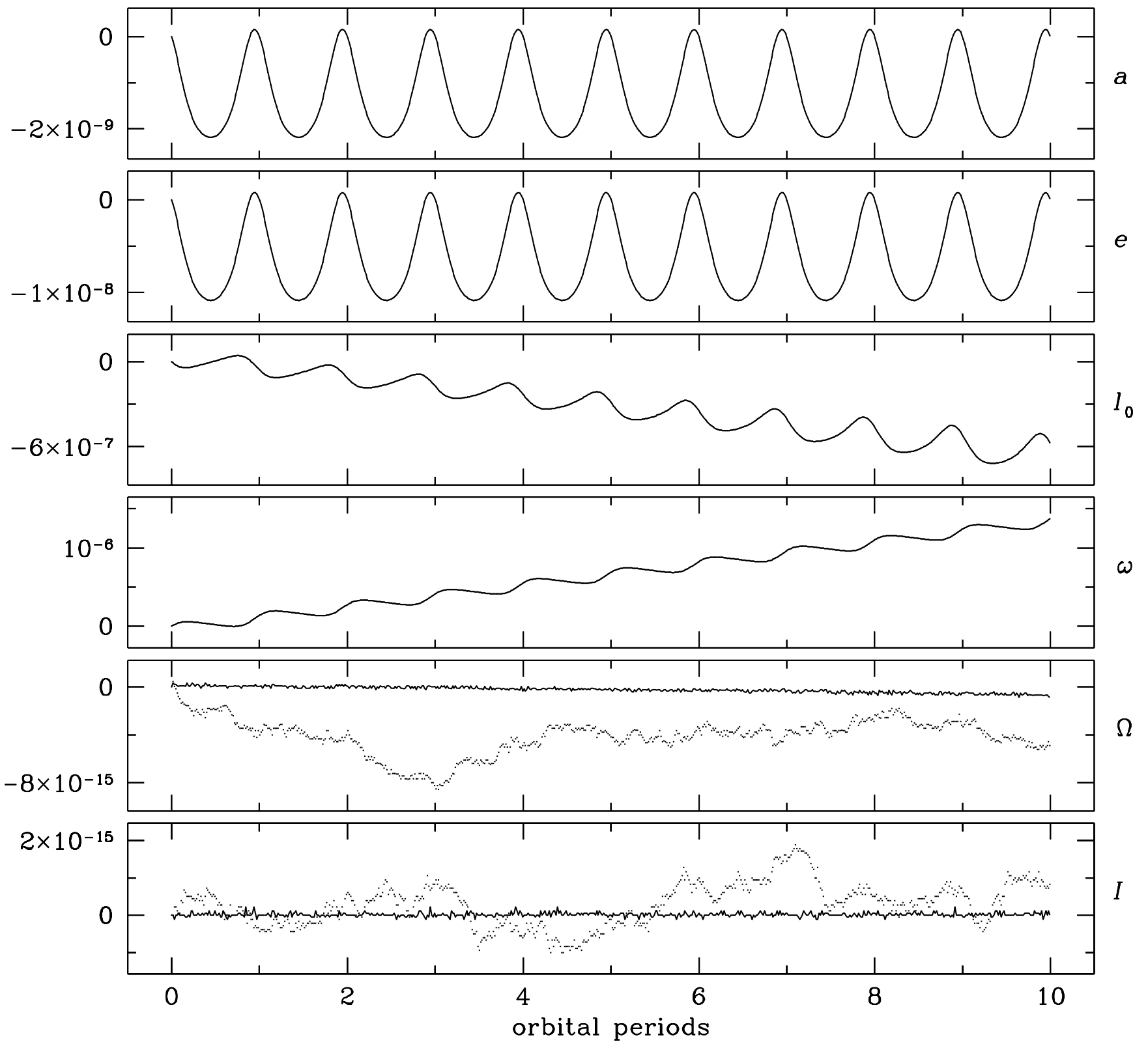

**Fig. 1** Errors in the orbital elements during 10 orbital periods of the gravitational two-body problem. Eight byte arithmetic was used with ordinary `Ruth_4` (dotted curves) and lattice `Ruth_4` with $m = 2^{62}$ (solid curves). The dotted and solid curves coincide in the top four panels. $\Omega$ and $I$ (bottom two panels) are more accurately conserved by the lattice method. In the exact solution, all six orbital elements are exactly conserved

In Fig. 1, there are small-amplitude oscillations in $a$, $e$, $\ell_0$ and $\omega$, the elements dominated by truncation error; such oscillations are expected in any symplectic integrator. In addition, the error in $\ell_0$ grows because the numerical scheme has a slightly incorrect period, and the error in $\omega$ increases because the numerical potential is slightly non-Keplerian so the orbit precesses. There is no reason to expect a symplectic algorithm to avoid these secular errors. (Of course, they are smaller with higher order SIAs.)

The characteristic oscillations in the elements do not occur if roundoff dominates the computational error. Fig. 2 shows integrations for the same

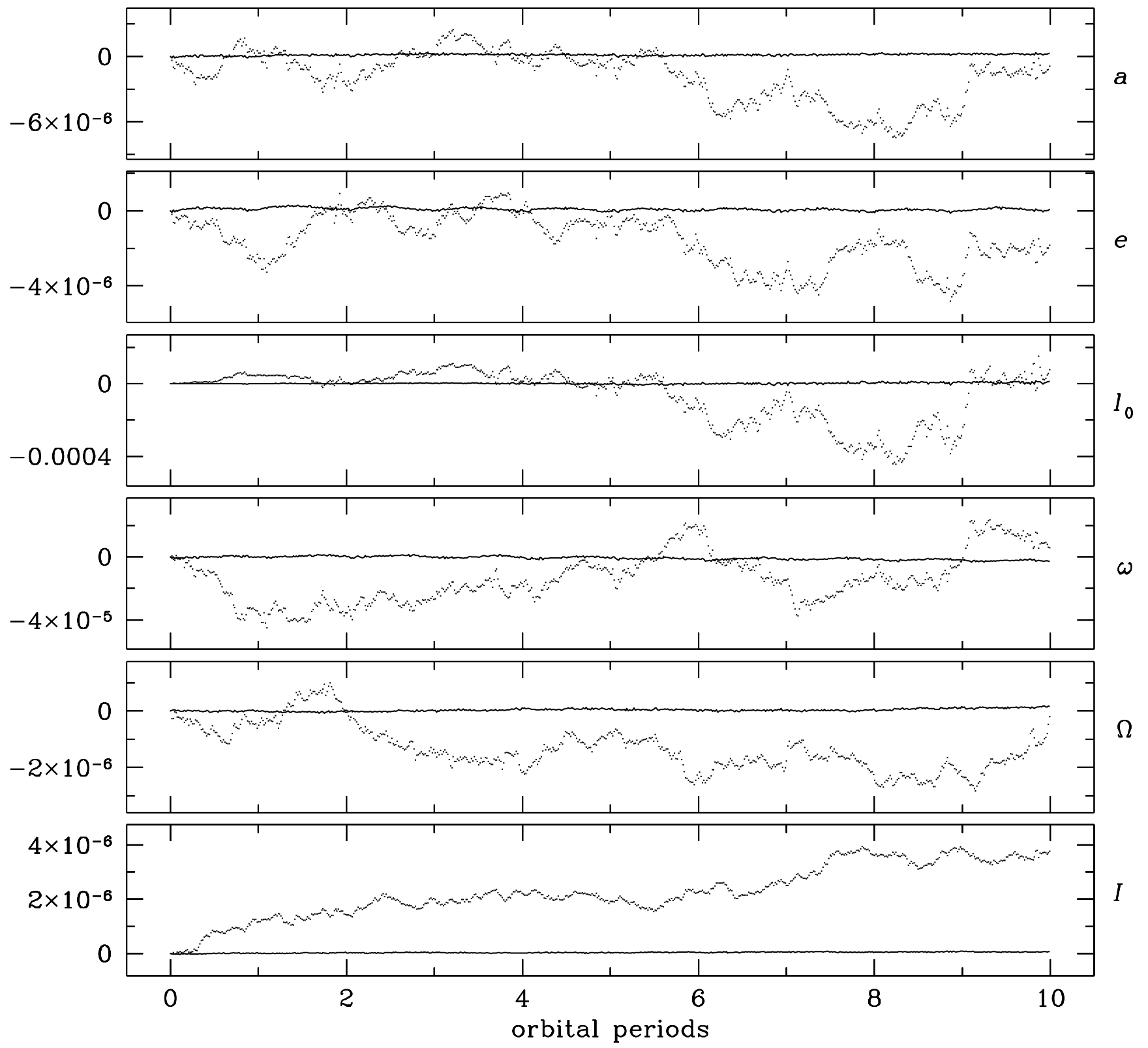

**Fig. 2** Integrations of the two-body problem with four byte arithmetic. Lattice `Ruth_4` with $m = 2^{30}$ (solid curves) yields more accurate results for all six elements

number of steps as in Fig. 1, this time using four byte (32 bit) arithmetic, i.e., half the precision. The error ratio is now $\mathcal{R}(24, 0.01, 4) \simeq 600$ so the lattice approach is expected to help. However, to avoid degrading the force and velocity components we require $m = 2^{24}/0.01 \sim 2^{31}$. When using four byte integers it is not practical to use $m > 2^{30}$ since this would risk integer overflow from additions, so $m = 2^{30}$ was used for the test shown in Fig. 2. Despite this, lattice `Ruth_4` conserves *all* the elements more accurately than ordinary `Ruth_4`. The improvement is by a factor of about 20 for $a$, 14 for $e$, 21 for $\ell_0$, 14 for $\omega$, 14 for $\Omega$, and 33 for $I$ (maximum errors in energy and angular momentum are each reduced by factors of about 20). The improvement is greater if the error ratio is larger: repeating the integration with $\Delta t = 0.002$ yields factors of about 36 for $a$, 51 for $e$, 34 for $\ell_0$, 107 for $\omega$, 22 for $\Omega$, and 42 for $I$ (and 36 for $E$ and $h$).

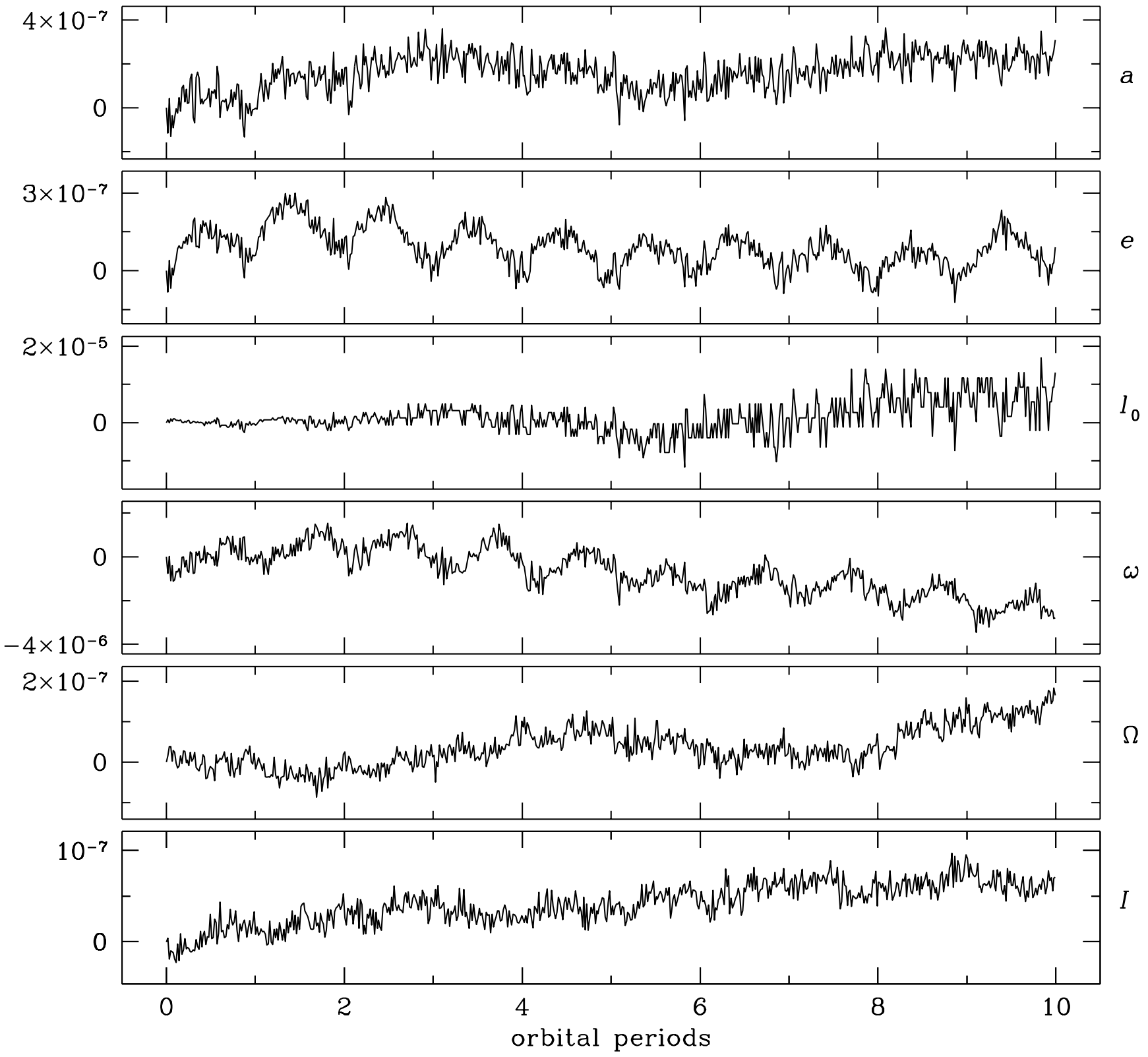

**Fig. 3** Expanded scale plots of the results of the lattice `Ruth_4` integration shown in Fig. 2. Oscillations typical of SIAs are evident

Beyond the reduction in the magnitude of the errors that is obvious in Fig. 2, an expanded plot of the lattice integration (Fig. 3) shows that exact symplecticity has restored the expected periodicity in the elements (see $e$ and $\omega$ especially). There appears to be an extra oscillation on a timescale of about five orbital periods and in general the curves are nowhere near as smooth as those in Fig. 1. The lack of smoothness is not surprising since to avoid roundoff error the lattice method introduces very small scale fluctuations in the Hamiltonian; these are not present in the `Ruth_4` algorithm itself. Note that the periodicity observed in Fig. 3 is *not* the oscillation due to truncation error that is intrinsic to `Ruth_4`—the amplitude of that is smaller by two orders of magnitude (compare Figs. 1 and 3).

## 5. Discussion

This paper has shown that the lattice `Ruth_4` algorithm outperforms ordinary `Ruth_4` when the computational error is dominated by roundoff ($\mathcal{R} \gg 1$). In integrations of the gravitational two-body problem with a time-step $\Delta t = 0.01$, the magnitude of the errors is reduced by a factor of order 20 (Fig. 2) and the character of the errors is more like what is found when truncation error dominates (compare Fig. 3 to the ordinary `Ruth_4` integrations in Figs. 1 and 2). The reduction of errors is more significant in going from ordinary `Ruth_4` to lattice `Ruth_4` than from a fourth-order (non-symplectic) Runge-Kutta integrator to ordinary `Ruth_4` (the experiment conducted by KYN).

The time-scale considered here is very short (10 orbital periods) but similar improvement occurs in much longer integrations. Using the lattice approach can clearly have noticeable qualitative effects on the dynamics only after a very long time. Studies that span millions of orbital periods may benefit significantly from the use of lattice SIAs.

Yoshida (1990) extended the Ruth class of SIAs to arbitrary (even) order; the lattice approach applies to all these integrators. More work is necessary to find high order SIAs whose efficiency compares well with the multi-step methods that are currently used in solar system integrations (e.g., Quinn, Tremaine & Duncan 1991). This is a challenging problem because some of the free parameters that could be used to increase order must be used to arrange symplecticity.

To force roundoff to dominate the `Ruth_4` integrations, four byte arithmetic was used. The factor by which the elements are conserved better by lattice `Ruth_4` increases as the time step is reduced, i.e., as roundoff becomes relatively more important. This suggests that the relative gain achieved by lattice methods is likely to be more significant for higher order integrators (for which $\mathcal{R} \gg 1$ even when using eight byte arithmetic). In long integrations requiring high order methods, such as simulations of the solar system, lattice SIAs may offer the best way to reduce the effects of numerical errors on dynamical evolution.

The $m = 2^{62}$ lattice integration shown in Fig. 1 was done on a Convex computer, which provides full eight byte integers in hardware. Unfortunately, most computers do not supply full length integers in hardware, i.e., the longest integers are not as long as the longest floating-point numbers. Integer arithmetic can be done using the longest floating-point numbers, which typically raises the maximum lattice size from $m = 2^{30}$ to $m = 2^{52}$. This is very valuable for studies of maps (e.g., ET) but it is usually not sufficient to benefit lattice SIAs (because of the requirement that $m\Delta t \geq 2^P$ to avoid degrading the force and velocity).

Lattice methods are useful for studies of the long-term evolution of Hamiltonian systems. For Hamiltonian maps, it is usually advantageous to use the lattice approach. For continuous Hamiltonian systems, lattice methods sig-nificantly reduce the errors in *high order* SIAs. This improvement is obtained without significantly compromising efficiency *provided* that full length integers are available in hardware. Thus, for a Hamiltonian problem requiring a high order integrator, and a computer with full length integers, a lattice SIA should be used.

**Acknowledgments.** It is a pleasure to thank Scott Tremaine for helpful advice and comments, and Moti Milgrom for his hospitality at the Weizmann Institute. This research was supported by a Weizmann Institute Exchange Fellowship and a grant from the Albert Einstein Centre for Theoretical Physics.